\documentclass [12pt]{article}
\usepackage{jheppub}
\usepackage{amsfonts}
\usepackage{graphicx}
\usepackage{amsmath,amssymb, amsfonts,color}
\usepackage{latexsym}
\usepackage{mathrsfs}
\usepackage{bbold}
\usepackage{tabularx}
\usepackage{color}
\usepackage{slashed}
\usepackage{cancel}
\usepackage{soul} %
\usepackage{setspace} %
\usepackage{comment}
\usepackage{booktabs}
\usepackage{multirow}
\usepackage{tikz-feynman}
\tikzfeynmanset{compat=1.0.0}
\usepackage{epsfig}
\usepackage{slashed}
\usepackage{epstopdf}
\usepackage{latexsym}
\usepackage{subcaption}
\usepackage{mdframed}
\usepackage{bm}
\usepackage{upgreek}

\usepackage{tikz}
\usetikzlibrary{decorations.pathmorphing}
\pgfdeclaredecoration{complete sines}{initial}
{
    \state{initial}[
        width=+0pt,
        next state=sine,
        persistent precomputation={\pgfmathsetmacro\matchinglength{
            \pgfdecoratedinputsegmentlength / int(\pgfdecoratedinputsegmentlength/\pgfdecorationsegmentlength)}
            \setlength{\pgfdecorationsegmentlength}{\matchinglength pt}
        }] {}
    \state{sine}[width=\pgfdecorationsegmentlength]{
        \pgfpathsine{\pgfpoint{0.25\pgfdecorationsegmentlength}{0.5\pgfdecorationsegmentamplitude}}
        \pgfpathcosine{\pgfpoint{0.25\pgfdecorationsegmentlength}{-0.5\pgfdecorationsegmentamplitude}}
        \pgfpathsine{\pgfpoint{0.25\pgfdecorationsegmentlength}{-0.5\pgfdecorationsegmentamplitude}}
        \pgfpathcosine{\pgfpoint{0.25\pgfdecorationsegmentlength}{0.5\pgfdecorationsegmentamplitude}}
}
    \state{final}{}
}

\usepackage{rotating}
\usepackage{bbm}
\usepackage{bbold}
\usepackage{enumitem}

\usepackage{empheq}
 \usepackage{caption}

\usepackage{amsthm}

\theoremstyle{definition}

\newtheorem {remark} {Remark}[section]

\definecolor{myblue}{rgb}{.8, .8, 1}

\definecolor{red}{rgb}{1,0,0}
\usetikzlibrary{positioning}
\tikzset{
    mybluenode/.style={
        draw=black, circle, minimum width=2cm, inner sep=0pt
        },
    myblacknode/.style={
        circle, inner sep=1pt, fill=black
        },
    }

\newcommand*\be{\beta}

\newcommand{\bea}{\begin{equation} \begin{aligned}} \newcommand{\eea}{\end{aligned} \end{equation}}
\def\be{\begin{equation}} \def\ee{\end{equation}} 
\def\beq{\begin{equation}} \def\eeq{\end{equation}}

\usepackage[english]{babel}
\usepackage{csquotes}
\MakeOuterQuote{"}

\makeatletter
\def\@fpheader{\ }
\makeatother

\title{\Large Disturbing news about the $d=2+\epsilon$ expansion II.\\Assessing the recombination scenario}
\author{Fabiana De Cesare$^a$}
\author{and Slava Rychkov$^{b,c}$}

\affiliation{$^a$INFN, Sezione di Torino, and Department of Physics, University of Turin,
Via P. Giuria 1, 10125, Turin, Italy}
\affiliation{$^b$Institut des Hautes \'Etudes Scientifiques, 91440 Bures-sur-Yvette, France}
\affiliation{$^c$Yukawa Institute for
	Theoretical Physics, Kyoto University,
	Kitashirakawa Oiwakecho, Sakyo-ku, Kyoto 606-8502, Japan}
\emailAdd{fabiana.decesare@unito.it}
\emailAdd{slava@ihes.fr}
	
\abstract{In \cite{paper1}, we revisited the $d=2+\epsilon$ expansion in the $O(N)$ Non-Linear Sigma Model (NLSM), emphasizing the existence of a protected operator which is a closed form with $N-1$ indices. The scaling dimension of this operator stays exactly equal to $N-1$, independently of $\epsilon$. Its existence is problematic for the identification of the NLSM fixed point in $d=2+\epsilon$ with the Wilson-Fisher fixed point family obtained by analytically continuing from near $d=4$, which does not possess such a protected operator. Multiplet recombination is one scenario discussed in \cite{paper1}, which could allow to connect the two families continuously (although not analytically). In this scenario, the protected dimension is lifted at some critical value of $\epsilon$, thanks to the short conformal multiplet of scaling dimension $N-1$ eating a long conformal multiplet of higher scaling dimension. In this followup work, we assess this scenario for the cases $N=3$ and $N=4$. We identify the lowest candidates for the long multiplet which could be eaten, and compute their one-loop anomalous dimensions. We find that at one loop, scaling dimensions of these candidates grow with $\epsilon$, while it should decrease down to $N$ for the recombination to occur. We conclude that multiplet recombination is unlikely.}

\begin{document}
	\maketitle
	\flushbottom
	
\section{Introduction}

The $d=2+\epsilon$ expansion is a time-honored technique for constructing perturbative conformal theories as UV fixed points of Non-Linear Sigma Models (NLSM) on positively curved manifolds, the most famous case being the $O(N)$ symmetric sigma-model with the target space $S^{N-1}$ \cite{Polyakov:1975rr,Brezin:1975sq,Bardeen:1976zh,Brezin:1976ap,Brezin:1976qa,Hikami:1977vr}. Until not so long ago it was believed that this $O(N)$ NLSM fixed point in $d=2+\epsilon$ coincides with the Wilson-Fisher (WF) fixed point analytically continued from $d=4-\epsilon$. We recently put this into question \cite{paper1}, following prior observations in \cite{Nahum_2015,Jones:2024ept}. The problem is that the $d=2+\epsilon$ fixed point has a protected operator---a closed form with $N-1$ indices, constructed as a pullback of the volume form on the target manifold. The dimension of this operator takes a protected value $N-1$, independently of $\epsilon$. Since there is no such protected operator in the Wilson-Fisher family at $d=4-\epsilon$, the two families cannot be connected by analytic continuation.

There is still a possibility that the two families can be connected continuously, although not analytically \cite{paper1}. As discussed in \cite{paper1}, good agreement between Wilson-Fisher results and the exponents of physical models in $d=3$ suggests that if the continuous connection happens, it should occur at $d=2+\epsilon_c<3$. The protected operator dimension is then lifted at $\epsilon=\epsilon_c$. In conformal field theory, the dimension of a protected operator which is a primary of a short multiplet $S$ of the conformal group can only be lifted if this multiplet recombines with extra states, turning into a long multiplet $L$. Schematically,
\beq
S\cup L' \to L\,.\label{eq:recomb}
\eeq
The recombined long multiplet $L$ can then move away from the protected scaling dimension, when $\epsilon$ is further increased. 

$L'$ in \eqref{eq:recomb} stands for a (long) multiplet supplying the missing states which complete $S$ into $L$. Quantum numbers of this multiplet are not arbitrary. Suppose the shortening condition for the short multiplet primary $\mathcal{O}$ has the form
\beq
\mathcal{D}\mathcal{O} = 0
\eeq
where $\mathcal{D}$ is a differential operator of degree $n$. The primary $\mathcal{O}'$ of the multiplet $L'$ should transform under the conformal group in exactly the same way as the l.h.s. of this shortening condition. In particular, the scaling dimension of $\mathcal{O}'$ should be $n$ units higher than that of $\mathcal{O}$. The Lorentz and global symmetry quantum numbers of $\mathcal{O}'$ are also fixed in terms of those of $\mathcal{O}$.

In our problem the protected operator $\mathcal{O}$ is a closed form with $N-1$ indices, of dimension $N-1$. Under the global $O(N)$ rotation it is a pseudoscalar. The operator $\mathcal{D}$ is the exterior derivative. Therefore, the Lorentz and global symmetry quantum numbers of $\mathcal{O}'$ should be:
\beq
\mathcal{O}'\ :\  \text{$N$-index form, $O(N)$ pseudoscalar},
\label{eq:qn}
\eeq 
while its scaling dimension should be 
\beq
\Delta_{\mathcal{O}'}=N\quad(\text{needed}).
\label{eq:needed}
\eeq 

We showed in \cite{paper1} that the lightest primary operators with Lorentz and global quantum numbers as in \eqref{eq:qn} have scaling dimension $\ge N+4+O(\epsilon)$, thus proving that recombination cannot occur for infinitesimal $\epsilon$. The goal of this followup work is to assess if it can happen at a finite value of $\epsilon$. For $N=3,4$, we will identify the lightest candidates for $\mathcal{O}'$ and we will compute their one-loop anomalous dimensions. Does the full $O(\epsilon)$ scaling dimension increase or decrease with $\epsilon$? If it decreases, for what value of $\epsilon$ does it cross the needed value \eqref{eq:needed}? 

The paper is structured as follows. In Section~\ref{sec:candidates}, we discuss how to identify candidate operators for the recombination. In Section~\ref{sec:mixing}, we explain how to solve the mixing problem and extract the anomalous dimensions of these candidates. Finally, in Section~\ref{sec:results}, we present our results for $N=3$ and $N=4$ and discuss their physical implications.

\section{Search for candidate operators}
\label{sec:candidates}
The recombination scenario described in the introduction requires the existence of an operator $\mathcal{O}'$ characterized by the quantum numbers in \eqref{eq:qn}, which should also be a primary. The lightest operator with quantum numbers \eqref{eq:qn} contains $N+2$ derivatives, of which $N$ antisymmetrized and 2 contracted. But this operator is a descendant \cite{paper1}. Since we need a primary, we consider the second-lightest class of operators with quantum numbers \eqref{eq:qn}, namely those with $N + 4$ derivatives,  $N$ antisymmetrized and 4 contracted. The number of such operators is significant, and it will be somewhat challenging to count and identify independent primary operators, which are the only ones of interest. 

We proceed as follows. We begin by briefly reviewing how to efficiently construct $O(N)$ pseudoscalar operators, referring to App. A of \cite{paper1} for further details. We then present a method to identify the subset of independent operators in a given Lorentz representation, and finally provide a procedure to isolate the primary operators among these independent ones.  We keep $N$ general and fixed in this section, and will give the results for $N=3,4$ in the next one.\footnote{\label{note:ev}Note that any operator with more than $k$ antisymmetrized indices vanishes in the limit $d=k$, i.e.~it will be an evanescent operator. However, by the rules of quantum field theory in non-integer $d$, such operators are considered nontrivial for generic $d$, e.g.~they have nontrivial correlation functions. The recombination phenomenon we are interested in, if it happens, may happen at any $d$. So when we look for recombination candidates, we work at generic $d$.}

\subsection{Identification of independent operators}
Our starting point is the $O(N)$ non-linear sigma model defined by the action
\begin{equation}
S=\frac{1}{2t}\int d^d x\,(\partial_\mu n^a)^2,
\label{eq:action}
\end{equation}
where the $n^a$ ($a=1,\dots,N$)  are real scalar fields constrained to live on the unit sphere,
\begin{equation}
n^a n^a = 1\,.
\label{eq:constraint}
\end{equation}
Composite operators in this theory are then built by taking products of $n$ fields and their derivatives. We are interested in $O(N)$ pseudoscalar operators, so we restrict our attention to operators in which all color indices are contracted using a number of Kronecker tensors and exactly one Levi-Civita tensor. Such operators can be generally written in terms of the following building blocks, obtained by contracting $O(N)$ indices either with the Kronecker delta or with the Levi-Civita epsilon:
\begin{equation}
\begin{aligned}
D_{\bar\mu^{(1)}\bar\mu^{(2)}} 
&= \partial_{\bar\mu^{(1)}} n^a \, \partial_{\bar\mu^{(2)}} n^a , \\
E_{\bar\mu^{(1)}\ldots \bar\mu^{(N)}} 
&= \epsilon_{a_1\ldots a_N}\,
   \partial_{\bar\mu^{(1)}} n^{a_1}\cdots
   \partial_{\bar\mu^{(N)}} n^{a_N}\,,
\end{aligned}
\label{eq:blocks}
\end{equation}
where the barred greek letters stand for sets of Lorentz indices: 
\be
\bar\mu^{(j)} = \{{\mu_1,\ldots,\mu_{k_j}}\},\quad \partial_{\bar\mu}=\partial_{\mu_1}\ldots \partial_{\mu_{k_j}}\,.
\ee
$O(N)$ pseudoscalar operators are then built by multiplying a number of such blocks, including exactly one $E$-block,\footnote{Recall that the product of two Levi–Civita tensors can be reduced to products of Kronecker deltas.} and contracting (some of) the Lorentz indices within blocks and among different blocks. 

By imposing the constraint that fields live on the sphere and by eliminating redundant operators proportional to the equations of motion, the number of independent operators can be significantly reduced. We refer to Appendix A of \cite{paper1} for a detailed derivation, and summarize here the main rules for $O(N)$ pseudoscalars:
\begin{enumerate}
	\item \label{rule1}
	Eliminate building blocks of the form $D_{\emptyset\bar\mu}$.
	\item 
	Eliminate building blocks containing contracted indices within a string of derivatives acting on $n^a$.
    \item 
	Restrict to operators with exactly one $E$-block ($k_E=1$) and a number of $D$-blocks $k_D$ satisfying  
    \be
    k_D\le(p-N+1)/2\,,
    \ee
    where $p$ is the total number of derivatives appearing in the operator. 
    \item \label{rule4}
    Discard all operators that vanish identically due to the antisymmetrization of open Lorentz indices.\footnote{We have in mind identities like $\partial_{[\mu}n^a \partial_{\nu]}n^a=0$, which are true in any $d$. As stressed in note \ref{note:ev}, evanescent operators which vanish only in integer $d$ are not discarded.}
\end{enumerate}
These rules substantially reduce the number of operators that must be considered, but they are not sufficient to isolate a basis of independent operators. In particular, there exist nontrivial relations among operators that are not captured by the criteria above. In our previous work, we looked at operators in the $N$-form Lorentz representation with a single pair of contracted derivatives ($p=N+2$). In that case, we were able to find ad hoc relations -- again derived from the constraint \eqref{eq:constraint} -- that allowed us to reduce the list obtained from rules \ref{rule1}–\ref{rule4}  to the actual set of independent operators. This was feasible because the result of rules \ref{rule1}–\ref{rule4} was a small set of six operators. In this work, we are instead interested in operators with two pairs of contracted derivatives ($p=N+4$). By contrast with the previous case, rules \ref{rule1}–\ref{rule4} now produce a much larger list (with more than 40 operators), making a similar case-to-case analysis impractical. An alternative strategy, already mentioned in \cite{paper1}, is to resolve the constraint \eqref{eq:constraint}, expressing the operators in terms of $N-1$ components $n^i=\sqrt{t}\,\pi^i$, with $i=1,\dots,N-1$, with the remaining component given by
\begin{equation}\label{eq:constsigma}
\sigma=n^N= \sqrt{1- t\ \pi^2}\,.
\end{equation}
When the operators are expressed in terms of $\pi$, determining whether they are dependent becomes a standard linear-algebra problem. However, when expressed in terms of the $\pi$ fields, operators become very lengthy nonlinear expressions, where the nonlinearity originates from the square root in $\sigma$ and its derivatives. This requires an additional step to extract a basis of independent operators.  Schematically, after using \eqref{eq:constsigma}, the operators take the form
\begin{equation}\label{eq:opspi1}
   O^a=\sum_{k} \frac{C^a_{\ k} \ {O}^{k}(\pi)}{(1-t\ \pi^2)^{l_k^a/2}}\,,
\end{equation}
where $C^a_{\ k}$ are real coefficients, ${O}^{ k}(\pi)$ are operators built from products of $\pi$ fields and their derivatives, and the $l^a_k$ are odd integers. Here the index $a$ runs over all operators obtained after applying rules \ref{rule1}–\ref{rule4}.
To properly extract linear dependencies among these operators, we factor out the largest power of the denominator.  Denoting by $l_{\rm max}$ the largest value of $l^a_k$ appearing in \eqref{eq:opspi1}, we can rewrite
\begin{equation}
\label{eq:opspi2}
O^a=\frac{1}{(1-t\,\pi^2)^{l_{\rm max}/2}}
\sum_k \widetilde{C}^a_{\ k}\,\widetilde{O}^k \, ,
\end{equation}
where the tildes indicate that, after extracting the common denominator, the coefficients and operators appearing in the linear combination have been redefined.
To count the number of independent operators, it is then sufficient to compute the rank $r$ of the coefficient matrix $\widetilde C$. 
A basis of independent operators is finally obtained by selecting the operators $O^a$ corresponding to a maximal set of linearly independent rows of $\widetilde C$. We will denote this basis by $B=\bigl\{\, O^1,\dots ,O^r\,\bigr\}$. As mentioned, we work at a generic $d$, and this basis is $d$-independent.

\subsection{Counting primaries}
In the previous section we presented a procedure to extract a basis of independent $O(N)$ pseudoscalar operators in a given Lorentz representation and with a given number of contracted derivatives.  For our purposes, we are interested in operators with $N$ antisymmetric open Lorentz indices and two pairs of contracted derivatives.  By applying the procedure described above, we obtain
\be\label{eq:basisop}
B_N=\bigl\{\, O^1_{\mu_1\cdots\mu_N},\,\dots,\,O^{r_N}_{\mu_1\cdots\mu_N}\,\bigr\}\,.
\ee
Now, operators in this list are linearly independent, but most of them turn out to be total derivatives, and will end up being descendants at the fixed point. Let $r_D$ be the dimension of the subspace of derivative operators. Then there are $r_P=r_N-r_D$ linearly independent operators which are not total derivatives, which will give rise to $r_P$ primaries. 

To determine $r_D$ and $r_P$, we proceed as follows. A moment’s thought shows that, in order to obtain (a linear combination of) operators in our list that is a total derivative, one must act with a derivative on an $O(N)$ pseudoscalar operator with $N-1$ antisymmetric open indices and two pairs of contracted derivatives, i.e.~with one fewer antisymmetrized derivative than the operators in $B_N$.\footnote{We note that this is the only possible way to obtain total derivatives in the $N$-form Lorentz representation because there are no lighter primary operators in this representation. Otherwise, one could also obtain a total derivative by acting with a box on such operators.} So we begin by counting linearly independent operators of the latter type. Applying the procedure described in the previous section, we get the following independent operators
\be
B_{N-1}
=\bigl\{\, O^1_{\mu_1\cdots\mu_{N-1}},\,\dots,\,O^{r_{N-1}}_{\mu_1\cdots\mu_{N-1}}\,\bigr\}\,.
\ee
We then take total derivatives of the operators in this basis and antisymmetrize the resulting index with the remaining open indices. This produces a set of total derivative operators in the $N$-form Lorentz representation,
\be \label{eq:descop}
\bigl\{\, D^1_{\mu_1\dots\mu_N},\,\dots,\,D^{r_{N-1}}_{\mu_1\dots\mu_N}\,\bigr\},
\ee
which are not necessarily linearly independent. To determine how many of them are independent, we re-express these total derivatives in the form $\eqref{eq:opspi2}$ and compute
the rank of the corresponding matrix $\widetilde{C}$. This rank is precisely $r_{D}$.

Let us replace \eqref{eq:basisop} by a new, equivalent, basis of operators, which includes linearly independent total derivatives as its first $r_{D}$ elements, while remaining $r_P$ operators are selected from the operators in \eqref{eq:basisop}, so that the total rank remains equal to $r_N$:
\be 	
\label{eq:finalbasis}
B_N=\bigl\{\, D^1_{\mu_1\dots\mu_N},\,\dots,\,D^{r_{D}}_{\mu_1\dots\mu_N}, O^1_{\mu_1\dots\mu_N},\,\dots,\,O^{r_P}_{\mu_1\dots\mu_N}\,\bigr\}\,.
\ee

By construction, operators $O^1,\ldots O^{r_P}$ are not total derivatives, and they will give rise to primaries at the fixed point.
To see how this happens, we will recall some basic facts about renormalization of composite operators in quantum field theory.
A generic set of operators $\{ \mathcal{O}^a \}$ having the same bare dimension at $\epsilon=0$ and same symmetry quantum numbers mix under the RG flow. This mixing can be encoded by the mixing matrix, which relates renormalized and bare operators. At one loop, which will be enough for our purposes here, this relation takes the form
\begin{equation}
	{\mathcal O}_R^a=(\delta^{a}_{\ b}+\delta Z^{a}_{\ b}){\mathcal O}^b\,,
\end{equation}
where $\delta Z$ is generically non-diagonal. Diagonalizing this matrix yields a set of eigenvalues $\delta_a$ and corresponding eigenvectors $\overline{\mathcal O}_a$, satisfying
\begin{equation}
	\overline{\mathcal O}_R^a=(1+\delta_{a})\overline{\mathcal O}^a\,.
\end{equation}
At the fixed point, these operators have definite scaling dimensions and therefore correspond either to primary operators or to descendants. 
 
 When we apply this procedure to our basis \eqref{eq:finalbasis}, the mixing matrix takes the following block-diagonal form:
\begin{equation}\label{eq:deltaZ}
\delta Z=
\begin{pmatrix}
A_{r_D\times r_D} &0\\
B_{r_P\times r_D}& C_{r_P\times r_P}\
\end{pmatrix}\,.
\end{equation}
This means that total derivative operators only mix among themselves. This is easy to understand. Total derivative operators are renormalized automatically once we renormalize the operator of which they are total derivative. So there cannot be any need to include a counterterm proportional to non-total-derivative operators. 

This implies that, among the eigenvectors, we can identify $r_{D}$ which are only written as linear combinations of the $D^i$. These will correspond to the descendants. The remaining $r_{P}$ eigenvectors involve linear combinations of $O$'s and $D$'s. These will be primaries.

In this paper, we will only compute anomalous dimensions of the primaries, not their full expressions. We will then compute the matrix $C$, whose eigenvalues determine these anomalous dimensions. We will not compute matrices $A$ and $B$. These would be useful for the full expressions for the primary operators, but, as mentioned, these are not relevant for our purposes.

\section{Solving the mixing problem}
\label{sec:mixing}

We now describe how to compute the anomalous dimension of the primary operators. To simplify the notation, we focus on the case of a single primary operator, i.e. $r_P=1$. We will see below that this is precisely what happens for $N=3,4$, see Sec.~\ref{sec:results}. 
With $r_P=1$, the basis in \eqref{eq:finalbasis} takes the form
\begin{equation}
\label{eq:finalbasis2}
B_N=\bigl\{\, D^1_{\mu_1\dots\mu_N},\,\dots,\,D^{r_{N}-1}_{\mu_1\dots\mu_N}, O_{\mu_1\dots\mu_N}\,\bigr\}\,,
\end{equation} 	
where $O$ is the only operator which is not a total derivative. In this setup, the matrix $C$ in \eqref{eq:deltaZ} reduces to a number, which determines the anomalous dimension of the unique primary operator. To compute it, it therefore suffices to renormalize the operator $O$ and isolate the counterterm proportional to the operator $O$ itself. In other words, this requires to compute the divergent part of correlation functions involving $O$, namely 
\begin{multline}
  \left.\langle n(x_1)\dots n(x_\ell) O(x)\rangle_\mathrm{NL} \right|_\mathrm{div} \\ =\delta_{O}\langle n(x_1)\dots n(x_\ell) O(x)\rangle_\mathrm{L}+\sum_{j=1}^{r_N-1}\delta_{O,D^{j}}\langle n(x_1)\dots n(x_\ell) D^j(x)\rangle_\mathrm{L}\,,
    	\label{eq:delta_O1PI}
    \end{multline}
    and to extract the value of $\delta_O$. Here we have left $O(N)$ indices implicit and we have denoted with L and NL the leading and the next-to-leading order contributions in the coupling $t$ of the correlation function.  Below, once we rescale $O$ and choose indices of $n$ fields appropriately, L will stand for order $t^0$ and NL for $t^1$.

To extract meaningful information, the number $\ell$ of elementary fields in the correlation functions \eqref{eq:delta_O1PI} will have to be chosen with care. We will come back later to this point. 

From the knowledge of $\delta_O$, we can compute the anomalous dimension of the primary simply taking 
\begin{equation}
    \gamma=\frac{d \ \delta_O}{d \log \mu}\,.
    \label{eq:gamma}
    \end{equation}
    In dimension regularization and minimal subtraction, the counterterm $\delta_O$ will take the form\footnote{Here and in the following we write $t$ for simplicity of notation, but this should be understood as the renormalized coupling $t_R$.}
      \begin{equation}\label{eq:count}
    \delta_O= \frac{c_O}{\epsilon}t+O(t^2)\,,
    \end{equation}
where $c_O$ is a real number, and the $\beta$-function reads
    \begin{align} \label{eq:beta}
\beta( t)&=\frac{d t}{d \log \mu}=\epsilon t+O(t^2)\,.
\end{align}
    Then, to compute the anomalous dimension we can differentiate by $\mu$ Eq.~\eqref{eq:count}, use Eq.~\eqref{eq:beta}, and evaluate it at the fixed point\footnote{We refer to \cite[Sec.4.1.3]{Henriksson:2022rnm} and \cite[App.~C]{paper1} for a review of the useful perturbative results for the NLSM fixed point. }
    \be t^*=\frac{2\pi}{N-2}\epsilon+O(\epsilon^2)\,.\label{eq:FP}
    \ee

We now explain how to compute the counterterm $\delta_O$.
First, we rewrite the operators and the action in \eqref{eq:action} in terms of the unconstrained fields $\pi$. This gives
\begin{equation}
S=\int d^dx\left(\frac12\partial_\mu\pi  ^i\partial^\mu \pi  ^i+\frac{t}{2}\frac{(\pi  ^i\partial_\mu \pi  ^i)^2}{1-t\pi  ^i\pi  ^i}\right)\,
\end{equation}
for the action and Eq.~\eqref{eq:opspi1} for the operators.  Expanding any operator in $t$, we get an infinite series of terms with a number of $\pi$ fields increasing in steps of $2$. 

 Let $m=M(O)$ be the smallest number of $\pi$'s occurring in the operator $O$. We rescale $O$ so that the leading term, which has exactly $m$ $\pi$'s, has no dependence on $t$. We then expand the rescaled operator $\widetilde{O}$ in $t$:\footnote{We here use a slightly different notation with respect to that of \cite{paper1}.}
\begin{equation}
\begin{aligned}
&O = t^{m/2}\,\widetilde{O}\, , &\widetilde{O} = \sum_{k=0}^{\infty} t^k\, \widetilde{O}_{m+2k}\, .\\
\end{aligned}
\end{equation}
Only $\widetilde{O}_m$ and $\widetilde{O}_{m+2}$ will appear in the one-loop computation below. They contain $m$ and $m+2$ fields $\pi$ respectively.

Similarly, we can expand the operators $D^j$. We will adapt their expansion to that of $O$. We assume that $M(D^j)\ge M(O)=m$, i.e. that all $D^j$ contain at least $m$ $\pi$ fields at leading order. We then write
\begin{equation}
\begin{aligned}
&D^j = t^{m/2}\,\widetilde{D}^j\, ,
\qquad
&\widetilde{D}^j = \sum_{k=0}^{\infty} t^k\, \widetilde{D}^j_{m+2k}\, ,\
\end{aligned}
\end{equation}
where $\widetilde{D}_{m+2k}^j$ contains $m+2k$  fields $\pi$. Note that if $M(D^j)>m$ then $\widetilde{D}_{m}^j=0$; this is allowed by our assumptions.

We know that $\widetilde{O}$ cannot be expressed as a linear combination of $\widetilde D^j$. Let us assume furthermore that this is realized already at the leading order, i.e.~that $\widetilde{O}_m$ cannot be expressed as a linear combination of $\widetilde D_m^j$. This holds in the cases we will consider, and it simplifies the discussion. 
Under these assumptions, to extract $\delta_O$ we will set $\ell=m$ in  Eq.~\eqref{eq:delta_O1PI}, and take the $n$ fields to lie along the $\pi$ directions (with uncontracted indices).

The L correlation functions appearing on the right-hand side of \eqref{eq:delta_O1PI} are order $t^0$ and they arise at tree level:
\begin{equation}\label{eq:leading}
\begin{aligned}
\langle \pi(x_1)\cdots\pi(x_m)\,\widetilde O(x)\rangle_{\mathrm L}
&=
\langle \pi(x_1)\cdots\pi(x_m)\,\widetilde O_m(x)\rangle_0 \,,\\
\langle \pi(x_1)\cdots\pi(x_m)\,\widetilde D^j(x)\rangle_{\mathrm L}
&=
\langle \pi(x_1)\cdots\pi(x_m)\,\widetilde D^j_m(x)\rangle_0 \,.
\end{aligned}
\end{equation}
Here $\langle\cdots\rangle_0$ denotes correlation functions computed in the free theory of the $\pi$ fields. 

Let us discuss next the divergent part of the NL correlation function appearing on the left-hand side of \eqref{eq:delta_O1PI}. This is of order $t^1$ and arises at one loop. It receives two distinct contributions. The first arises from one-particle reducible (1PR) diagrams and is canceled by the wave-function renormalization of the elementary $\pi$ fields:
\begin{equation}
\left.
\langle \pi(x_1)\cdots\pi(x_m)\,\widetilde O(x)\rangle_{\mathrm{NL,\,1PR}}
\right|_{\mathrm{div}}
=
\frac{m}{2}\,\delta_\pi\,
\langle \pi(x_1)\cdots\pi(x_m)\,\widetilde O_m(x)\rangle_0 \,,
\end{equation}
where $\delta_\pi$ is the wave-function renormalization constant of $\pi$, which at one loop is 
\begin{equation}
	\label{eq:delta-pi}
	\delta_\pi=\frac{ t}{2\pi\epsilon}\,,
\end{equation}
independently of $N$ \cite[App.~C]{paper1}. The second contribution comes instead from the one-particle irreducible (1PI) diagrams and reads
\begin{multline}\label{eq:1PI}
\left.
\langle \pi(x_1)\cdots\pi(x_m)\,\widetilde O(x)\rangle_{\mathrm{NL,\,1PI}}
\right|_{\mathrm{div}}
\\=
\langle \pi(x_1)\cdots\pi(x_m)\,\widetilde O_m(x)\rangle_1
+
t\,\langle \pi(x_1)\cdots\pi(x_m)\,\widetilde O_{m+2}(x)\rangle_0 \,,
\end{multline}
where $\langle\cdots\rangle_1$ denotes 1PI correlation functions with a single insertion of the leading interaction vertex
\begin{equation}
\frac{t}{2}\int \mathrm d y\,(\pi^i\partial_\mu\pi^i)^2 \,.
\end{equation}
We now describe how to evaluate the two terms in Eq.~\eqref{eq:1PI}.  By renormalizability, both of these divergent pieces have to be expressible in terms of correlators of the lower-order components of the considered set of operators:
\begin{equation}\label{eq:1PIbis}
\begin{aligned}
&\langle \pi(x_1)\cdots\pi(x_m)\,\widetilde O_m(x)\rangle_1\\
&\hspace{1cm}=
\delta_O^{\mathrm A}\,
\langle \pi(x_1)\cdots\pi(x_m)\,\widetilde O_m(x)\rangle_0
+
\sum_j \delta^{\mathrm A}_{OD^j}\,
\langle \pi(x_1)\cdots\pi(x_m)\,\widetilde D^j_m(x)\rangle_0 \,,
\\
&\langle \pi(x_1)\cdots\pi(x_m)\,\widetilde O_{m+2}(x)\rangle_0\\
&\hspace{1cm}=
\delta_O^{\mathrm B}\,
\langle \pi(x_1)\cdots\pi(x_m)\,\widetilde O_m(x)\rangle_0
+
\sum_j \delta^{\mathrm B}_{OD^j}\,
\langle \pi(x_1)\cdots\pi(x_m)\,\widetilde D^j_m(x)\rangle_0 \,.
\end{aligned}
\end{equation}
 We need to determine $\delta_O^{\mathrm A}$ and $\delta_O^{\mathrm B}$, since they contribute to $\delta_O$, as follows:
\begin{equation}
	\delta_{O}
	=
	\frac{m}{2}\,\delta_\pi
	+
	\delta_O^{\mathrm A}
	+
	\delta_O^{\mathrm B} \,.
\end{equation}
On the other hand, coefficients $\delta^{\mathrm A}_{OD^j}$, $\delta^{\mathrm B}_{OD^j}$ contribute to the $B$ part of the mixing matrix \eqref{eq:deltaZ}. As mentioned, we will not need to compute them.

$\delta_O^{\mathrm A}$ can be evaluated using different methods. The standard approach consists of a systematic computation of momentum-space Feynman diagrams.\footnote{With this approach, there exists a choice of kinematics which is particularly convenient, which corresponds to setting the momentum of the operator $\widetilde{O}$ to zero \cite{Henriksson:2025vyi}. In this way, total derivatives decouple, making it easier to isolate the counterterm $\delta_O$.}  In our calculations we used an alternative technique relying on the position-space operator product expansion (OPE), see ~\cite{Cardy-book,Hogervorst:2015akt} for a review. 
 A practical description of this method in the present context is provided in App.~\ref{app:OPE}. Some of the results were double-checked via Feynman diagrams.
 
The computation of $\delta_O^B$ is simpler. As explained above, $\widetilde O_{m+2}$ is composed of $m+2$ $\ \pi$ fields. Of these, $m$ are Wick contracted with the external $\pi$ fields, while the remaining two are contracted with each other, giving rise to a tadpole, which can be evaluated straightforwardly.\footnote{
Evaluating the tadpoles, we find that all $\delta^{\mathrm B}_{O D^j}$ in Eq.~\eqref{eq:1PIbis} vanish. In other words tadpoles do not contribute to the operator mixing. This seems to be a general feature that we also observed in other cases, and it may have a deeper explanation that we did not explore in this work. On the other hand we did observe that the other mixing coefficients $\delta^{\mathrm A}_{O D^j}$ are generically nonzero.
}

\section{Results}\label{sec:results}
Let us now present the results that we obtained in the case $N=3,4$ following the procedure explained in the previous sections.  Since the analysis is technically similar in the two cases and many details have already been discussed, we will be brief.

\subsection{$N=3$}
We find 8 independent operators that are $O(3)$ pseudoscalars with three antisymmetric Lorentz indices and two pairs of contracted indices. 
Among these, there is only one primary operator. We are able to identify a basis of operators of the kind \eqref{eq:basisop}, corresponding to the following operators:
\begin{equation}
\begin{aligned}
&D^1_{\mu_1\mu_2\mu_3}=\mathcal{A}\circ\partial_{\mu_1}\left(E_{\{\kappa\nu\mu_2\},\{\kappa\nu\},\{\mu_3\}}\right)\,,
\\&D^2_{\mu_1\mu_2\mu_3}=\mathcal{A}\circ\partial_{\mu_1}\left(E_{\{\kappa\mu_2\},\{\kappa\nu\},\{\nu\mu_3\}}\right),
\\&D^3_{\mu_1\mu_2\mu_3}=\mathcal{A}\circ\partial_{\mu_1}\left(D_{\{\nu\mu_2\},\{\kappa\mu_3\}}E_{\{\kappa\},\{\nu\},\{\}}\right),
\\&D^4_{\mu_1\mu_2\mu_3}=\mathcal{A}\circ\partial_{\mu_1}\left(D_{\{\kappa\nu\mu_2\},\{\nu\}}E_{\{\kappa\},\{\mu_3\},\{\}}\right),
\\&D^5_{\mu_1\mu_2\mu_3}=\mathcal{A}\circ\partial_{\mu_1}\left(D_{\{\kappa\nu\},\{\nu\mu_2\}}E_{\{\kappa\},\{\mu_3\},\{\}}\right),
\\&D^6_{\mu_1\mu_2\mu_3}=\mathcal{A}\circ\partial_{\mu_1}\left(D_{\{\nu\mu_2\},\{\mu_3\}}E_{\{\kappa\},\{\kappa\nu\},\{\}}\right),
\\&D^7_{\mu_1\mu_2\mu_3}=\mathcal{A}\circ\partial_{\mu_1}\left(D_{\{\kappa\mu_2\},\{\nu\}}E_{\{\mu_3\nu\},\{\kappa\},\{\}}\right),
\\&O_{\mu_1\mu_2\mu_3}=\mathcal{A}\circ D_{\{\nu\mu_1\},\{\kappa\nu\mu_2\}}E_{\{\},\{\kappa\},\{\mu_3\}}\,,
\end{aligned}
\end{equation}
where $\mathcal{A}\circ$ denotes antisymmetrization of the indices $\mu_1\dots \mu_3$. The first 7 operators are total derivatives and the last operator contains the primary at the fixed point. 

\begin{figure}[h]  
	\centering
	\includegraphics[height=0.31\textwidth]{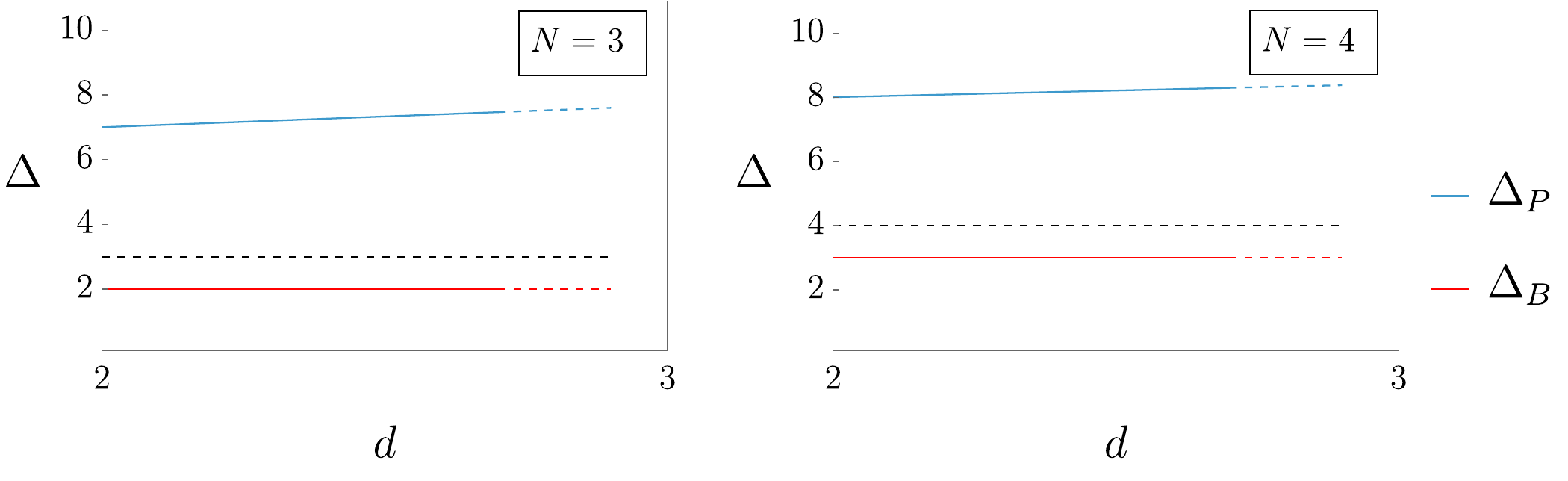}
	\caption{Scaling dimension of the lightest primary operator ($P$) in the $N$-form Lorentz representation in $d=2+\epsilon$ as a function of $d$ for $N=3$ (left) and $N=4$ (right). For recombination to occur, these dimensions must reach $\Delta=N$ (dashed line) in the interval $2<d<3$, which is strongly disfavored by our results. The scaling dimension of the protected operator ($B$) is also shown; it remains protected at $\Delta = N-1$.}
	\label{fig:mainresult}
\end{figure}

The leading term of $O$, $\widetilde O_0$, contains $m=4$ $\pi$ fields, while the leading terms of the total derivatives contain $\ge 4$ $\pi$'s. Moreover, $\widetilde O_0$ cannot be written as a linear combination of $\widetilde D^j_0$. So the assumptions of Section \ref{sec:mixing} are satisfied, and the anomalous dimension of $O$ can be computed as described there. Specifically, we obtain
\begin{equation}
\delta_O^\mathrm{A}=-\frac{2t}{3\pi\epsilon}\,,\quad \delta_O^\mathrm{B}=-\frac{t}{\pi\epsilon}\,,
\end{equation}
and as a consequence 
\begin{equation}
\gamma = -\frac{2t}{3\pi}\,.
\end{equation}
The classical scaling dimension of the primary is $\Delta_0 = 7 + 2\epsilon$, as it is composed at leading order of 7 derivatives and four $\pi$ fields, which have classical dimension equal to $\epsilon/2$. Evaluating the anomalous dimension at the fixed point \eqref{eq:FP}, we get
\begin{equation}
\Delta=7+\frac{2 \epsilon}{3}\,.
\end{equation}
As discussed in \cite{paper1} and in the introduction, for the recombination scenario to occur, we would need the dimension of this primary to decrease from 7 to 3 in the interval $0<\epsilon<1$. Our result instead shows that, at one loop, it grows with $\epsilon$. We conclude that recombination is unlikely to occur. Hence, the protected operator cannot be lifted in the interval of interest. Our result is illustrated in Fig.~\ref{fig:mainresult}.

\begin{remark}
	\label{rem:add}
	It's also interesting to probe the recombination scenario from the $d=4-\epsilon$ side. For $N=3$, the lightest operator in the $4-\epsilon$ expansion with the same Lorentz and global quantum numbers as the protected operator $B$ in $d=2+\epsilon$ expansion is \cite{paper1}
	\begin{equation}
		\label{eq:Btilde}
		\widetilde{B}_{\mu_1\mu_2}=\epsilon_{a_1a_2a_3}\partial_{[\mu_1}\phi^{a_1}\partial_{\mu_{2}]}\phi^{a_{2}} \phi^{a_3}\,.
	\end{equation}
	We can study the dimension of this operator as a function of $d$. We expect two possibilities.
Either 1) the dimension of $\tilde B$ approaches 2 the dimension of $B$ as $d\to 2$, or 2) it crosses 2 at some $d=d_c>2$. 
The second possibility would allow recombination at $d=d_c$. The first possibility would correspond to the scenario in \cite[Fig.5]{paper1} that the O(3) WF and NLSM fixed points are two distinct families of CFTs which however share the main scaling dimensions in the limit $d\to 2$.\footnote{The spectrum of O(3) NLSM at $d=2$ will be a subset of the limiting spectrum of the O(3) WF. The situation here will be analogous to that recently discussed by Zan for the Ising model \cite{Zan:2026oyb}, where the $d\to 2$ limiting spectrum is richer than the spectrum of the $d=2$ theory, to allow for the multiplet recombination as one moves away from $d=2$.} 

Recently, Refs.~\cite{Henriksson:2025hwi,Henriksson:2025vyi} studied scaling dimensions of many operators in the Wilson-Fisher fixed points in $4-\epsilon$ expansion, at five loops. Their results are exhaustive up to bare scaling dimension $6+O(\epsilon)$ and up to 2 Lorentz indices, and thus the five-loop anomalous dimension of \eqref{eq:Btilde} is available, see Eq.(15) of \cite{Henriksson:2025hwi}.\footnote{We thank J.~Henriksson for drawing our attention that there was a misprinting in this equation in the first arxiv version; here we are using the second arxiv version of \cite{Henriksson:2025hwi}.}

In Fig.~\ref{fig:Btilde}, we plot their result as a function of $2<d<4$, after Padé approximation. We see that the scaling dimension of $\widetilde{B}$ remains above 2 throughout most of the interval $2<d<4$, and approaches 2 at $d\approx 2.1$. This could in principle be interpreted as recombination at $d\approx 2.1$, but we find this unlikely: so close to $d=2$ our perturbative results precluding recombination should be trustworthy. The more natural interpretation of Fig.~\ref{fig:Btilde} is rather as an imperfect image of exact joining at $d=2$, imperfect since an approximate, five-loop result from the $d=4-\epsilon$ side is being used. In the future, this can be checked further by including more loops and seeing if the intersection point moves even closer to $d=2$.

\begin{figure} 
	\centering
	\qquad\includegraphics[width=0.7\textwidth]{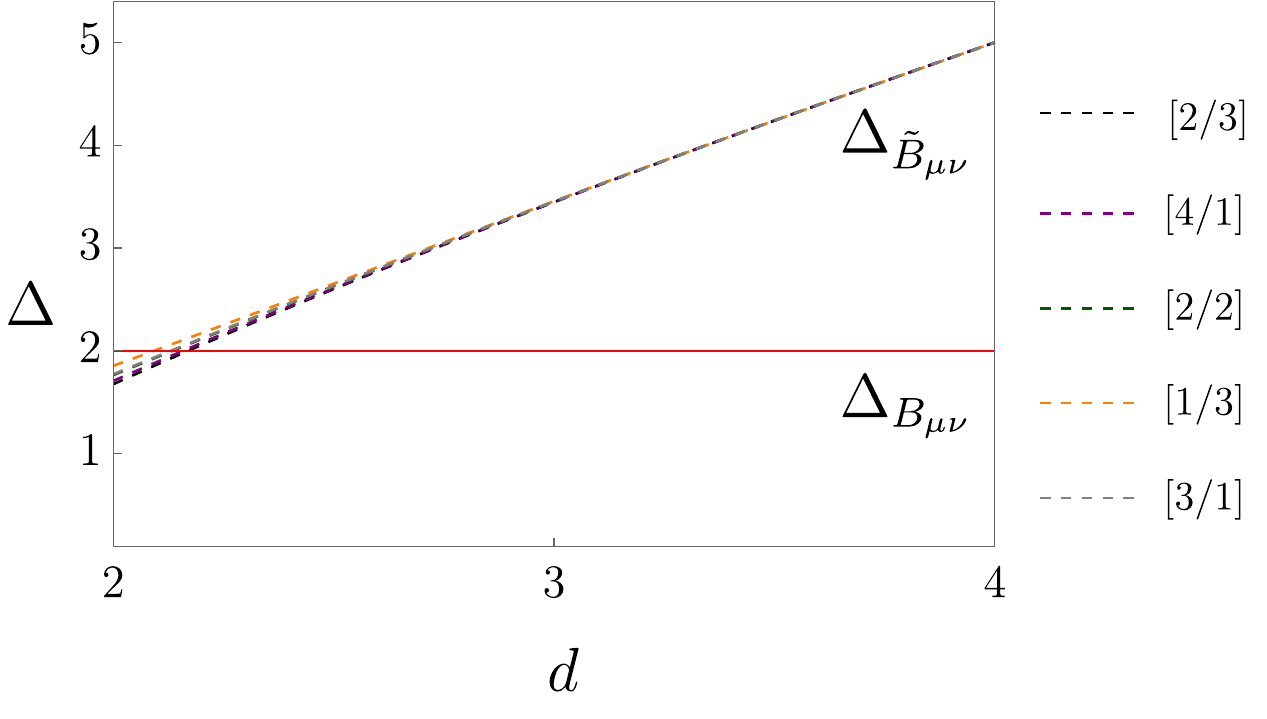} 
	\caption{Scaling dimension of the operator $\widetilde{B}$ as a function of $d$ (dashed lines), obtained via a Padé approximation of the results of \cite{Henriksson:2025hwi,Henriksson:2025vyi}, with the use of  Padé$_{m,n}$ approximants. For this test, we show all approximants with $m+n=4$ and 5 ($m,n\ge 1$), discarding those which have poles  in the range $2<d<4$. The Pade-approximated scaling dimension 
remains safely above that of the protected operator $B$ for most of the interval, and reaches $\Delta=2$ for $d$ close to 2. We view this numerical evidence as supporting the fact that the NLSM and the WF CFT only coincide at $d=2$.  }	
	\label{fig:Btilde}
\end{figure}
\end{remark}
 
 \subsection{$N=4$}
The analysis for $N=4$ proceeds along the same lines. In this case, we find 10 independent $O(4)$ pseudoscalar operators with four antisymmetric Lorentz indices and two pairs of contracted indices, among which a single primary operator is present. A basis of the form~\eqref{eq:basisop} is now given by
\begin{equation}
\begin{aligned}
&D^1_{\mu_1\mu_2\mu_3\mu_4}=\mathcal{A}\circ\partial_{\mu_1}\left(E_{\{\kappa\nu\mu_2\},\{\kappa\nu\mu_3\},\{\kappa\nu\},\{\mu_4\}}\right)\,,
\\&D^2_{\mu_1\mu_2\mu_3\mu_4}=\mathcal{A}\circ\partial_{\mu_1}\left(E_{\{\kappa\nu\mu_2\},\{\nu\mu_3\},\{\kappa\mu_4\},\{\}}\right),
\\&D^3_{\mu_1\mu_2\mu_3\mu_4}=\mathcal{A}\circ\partial_{\mu_1}\left(E_{\{\kappa\mu_2\},\{\nu\mu_3\},\{\kappa\mu_4\},\{\nu\}}\right),
\\&D^4_{\mu_1\mu_2\mu_3\mu_4}=\mathcal{A}\circ\partial_{\mu_1}\left(D_{\{\nu\mu_2\},\{\mu_3\}}E_{\{\kappa\mu_4\},\{\kappa\},\{\nu\},\{\}}\right),
\\&D^4_{\mu_1\mu_2\mu_3\mu_4}=\mathcal{A}\circ\partial_{\mu_1}\left(D_{\{\kappa\nu\mu_2\},\{\nu\}}E_{\{\mu_3\},\{\mu_4\},\{\kappa\},\{\}}\right),
\\&D^5_{\mu_1\mu_2\mu_3\mu_4}=\mathcal{A}\circ\partial_{\mu_1}\left(D_{\{\mu_2\},\{\nu\mu_3\}}E_{\{\mu_4\kappa\},\{\kappa\},\{\nu\},\{\}}\right),
\\&D^6_{\mu_1\mu_2\mu_3\mu_4}=\mathcal{A}\circ\partial_{\mu_1}\left(D_{\{\kappa\nu\mu_2\},\{\nu\}}E_{\{\mu_3\},\{\mu_4\},\{\}}\right),
\\&D^7_{\mu_1\mu_2\mu_3\mu_4}=\mathcal{A}\circ\partial_{\mu_1}\left(D_{\{\nu\mu_2\},\{\kappa\nu\}}E_{\{\mu_3\},\{\mu_4\},\{\}}\right),
\\&D^8_{\mu_1\mu_2\mu_3\mu_4}=\mathcal{A}\circ\partial_{\mu_1}\left(D_{\{\nu\mu_2\},\{\mu_3\}}E_{\{\kappa\nu\},\{\mu_4\},\{\kappa\},\{\}}\right),
\\&D^9_{\mu_1\mu_2\mu_3\mu_4}=\mathcal{A}\circ\partial_{\mu_1}\left(D_{\{\nu\kappa\},\{\mu_2\}}E_{\{\mu_3\nu\},\{\mu_4\},\{\kappa\},\{\}}\right),
\\&O^1_{\mu_1\mu_2\mu_3\mu_4}=\mathcal{A}\circ
D_{\{\nu\mu_1\},\{\kappa\nu\mu_2\}}
E_{\{\},\{\kappa\},\{\mu_3\},\{\mu_4\}}\,,
\end{aligned}
\end{equation}
where $\mathcal{A}\circ$ antisymmetrizes the indices $\mu_1\dots \mu_4$. The first nine operators are total derivatives and the last operator is the one containing the primary. 
Assumptions of Section \ref{sec:mixing} are satisfied with $m=5$. The counterterms and the anomalous dimension are given by
\begin{equation}
\delta_O^\mathrm{A}=-\frac{25t}{12\pi\epsilon}\,,\quad \delta_O^\mathrm{B}=-\frac{5t}{4\pi\epsilon}\,,\quad \gamma = -\frac{25t}{12\pi}\,.
\end{equation}
The classical scaling dimension of the primary is $\Delta_0 = 8 + 5\epsilon/2$, as it is composed of eight derivatives and five $\pi$ fields at leading order. Evaluating the anomalous dimension at the fixed point \eqref{eq:FP} 
then gives
\begin{equation}
\Delta = 8 + \frac{5}{12}\,\epsilon\,.
\end{equation}
In this case, the recombination scenario would require the dimension of the primary to approach 4 for some $0 < \epsilon < 1$. As in the $N=3$ case, our findings strongly disfavor this possibility, see Fig.~\ref{fig:mainresult}. 

\begin{remark}
	\label{rem:add1}
	
	In the future it would be interesting to provide additional evidence using considerations from the Wilson-Fisher fixed point, as we did for $N=3$ in Remark \ref{rem:add}. For that, we would need to consider the $N=4$ analogue of the operator \eqref{eq:Btilde}
	\begin{equation}
		\label{eq:Btilde1}
		\widetilde{B}_{\mu_1\mu_2 \mu_3}=\epsilon_{a_1a_2a_3 a_4}\partial_{[\mu_1}\phi^{a_1}\partial_{\mu_{2}}\phi^{a_{2}} \partial_{\mu_{3}]}\phi^{a_{3}}\phi^{a_4}\,,
	\end{equation}
	study its dimension in $d=4-\epsilon$ expansion, and compare it with the dimension of the protected operator, which is now equal to 3. This check is not currently available, because the five-loop results of \cite{Henriksson:2025hwi,Henriksson:2025vyi} do not cover operators with 3 Lorentz indices.
\end{remark}

\section{Conclusions}

In this work, we have extended the analysis initiated in \cite{paper1}, regarding the presence of a protected operator in the NLSM $O(N)$ CFT which is absent in the WF $O(N)$ CFT, preventing them from being analytically connected in the interval $2<d<4$. To explore a possible reconciliation of the statement that the two CFTs are connected—albeit only continuously—we investigated the possibility of multiplet recombination for the protected operator in the NLSM $O(N)$ CFT in $d=2+\epsilon$. 
 This scenario was already discussed in \cite{paper1}. However, the lack of quantitative testing made it impossible to draw strong conclusions. The analysis presented here allows us to better assess its likelihood.

We focused on the cases $N=3$ and $N=4$, for which we 
found the lightest primary operators that could be involved in recombination, namely pseudoscalar operators in the Lorentz $N$-form with dimension $N+4 +\mathcal{O}(\epsilon)$. We then solved the mixing problem under renormalization and computed their one-loop anomalous dimensions. In both cases, we found positive anomalous dimensions, providing strong evidence that the multiplet recombination scenario is not realized, as this would require the dimension to decrease down to $N$.  Barring higher-loop corrections or non-perturbative effects which could in principle modify this conclusion, our results suggest the absence of a continuous connection between the two families of CFTs.

 This result allows us to address the main question raised in \cite{paper1}, at least for $N=3$ and $N=4$. Indeed, in these cases, once multiplet recombination is ruled out, the only possible conclusion is that the two CFTs are distinct theories. We note that \cite{paper1} mentions another scenario, in which the two theories match only at $d=3$, but are distinct for non-integer values of dimensions. However, this scenario remains viable---though admittedly far-fetched---only for $N>4$, as the protected operator vanishes at this specific value of the dimension, allowing the two theories to coincide. By contrast, for 
$N=3$ and $N=4$, the protected operator is non-vanishing at $d=3$, and this scenario is therefore excluded.  
We then conclude that in these cases the NLSM $O(N)$ CFT is most likely a distinct theory from the WF $O(N)$ CFT throughout the interval $2<d<4$. 
 As explained in \cite{paper1}, the scenario of distinct theories is further supported by the presence of a singlet scalar operator approaching marginality in the NLSM $O(N)$ CFT at some  critical value $d_*>2$. Such behavior suggests the disappearance of the NLSM $O(N)$ CFT through a mechanism of merger and annihilation at $d=d_*$. 
It would be interesting to test further whether $d_*<3$ or $d_*>3$. In the latter case this  CFT survives as a real CFT in physical dimensions $d=3$. This could be investigated perturbatively, by computing the dimensions of candidate marginal operators at higher orders in $\epsilon$, or non-perturbatively, by using the conformal bootstrap  and looking for three-dimensional CFTs with $O(N)$ symmetry that are distinct from the WF $O(N)$ CFT. 
In particular, one could analyze constraints on correlation functions with the protected operator taken as an external operator as this, being present only in the NLSM  $O(N)$ CFT, would provide a clear way to differentiate between the two theories. We note that in the cases $N=3$ and $N=4$, the analysis is simplified by the fact that, after Hodge dualization, the protected operator becomes a conserved current and a pseudoscalar, respectively, which are simpler objects to study. Of course, there remains the possibility that such a search may not find any consistent theory, if the merger occurs before $d=3$.

Finally, it would be interesting to extend this analysis to larger values of $N$. Since identifying the set of independent primary operators—and especially computing their anomalous dimensions—becomes increasingly difficult as $N$ grows, it would be valuable to develop a method to compute a general result, or at least formulate an educated guess for the generalization of our one-loop findings.

For further future directions we refer to \cite{paper1}, where a comprehensive list of open questions was outlined. We believe that the topic remains highly relevant, and that the results of the present work provide additional motivation for studying it.
\acknowledgments
    
 FDC thanks Stéphane Bajeot, Johan Henriksson, and Marten Reehorst for useful discussions. FDC is supported by the Italian Ministry of University and Research (MUR) under the FIS
grant BootBeyond (CUP: D53C24005470001), and by the
INFN “Iniziativa Specifica” ST$\&$FI. The authors thank the Yukawa Institute for Theoretical Physics at Kyoto University, where discussions during the “Progress of Theoretical Bootstrap” program were useful in completing this work.

\appendix
\section{Anomalous dimensions from the OPE in the NLSM}
\label{app:OPE}
 In this appendix we describe how to compute the one-loop anomalous dimensions of composite operators in position space from the OPE. This method is well known for the Wilson-Fisher renormalization in $d=4-\epsilon$ \cite{Cardy:1996xt}. It is convenient to simplify combinatorics when complicated composite operators are involved \cite{Hogervorst:2014rta}, which is also our case. Here we will describe how to adapt it to the NLSM. Consider the correlation function
\begin{equation}
\langle O(0)\,\dots \rangle_1
=
-\frac{t}{2}\langle O(0)\int \mathrm d^d y\,V(y) \,\dots \rangle \,, 
\end{equation}
where $O$ is a local operator containing $m$ $\pi$ fields and $p$ derivatives and $V$ is the leading interaction term
\begin{equation}
V(y) =(\pi^i(y)\partial_\nu\pi^i(y))^2\,.
\end{equation}
The basic idea of this method relies on replacing the product between $O$ and $V$ with their OPE expansion, focusing only on the terms which give a non-vanishing pole in $\epsilon$, when integrated in $y$ in dimensional regularization. 
The procedure is as follows:
\begin{enumerate}
\item Perform Wick contractions between two of the four fields in $V$ and two of the $m$ fields in $O$, using the two-point function
\begin{equation}\label{eq:twopoint}
\left\langle \pi^i(y)\pi^j(x)\right\rangle
=
\frac{\Gamma\!\left(\frac{d}{2}\right)}{2(d-2)\pi^{\frac{d}{2}}|x-y|^{d-2}}\,
\delta^{ij},
\end{equation}
and its derivatives.  
Specifically, let $s$ denote the number of derivatives acting on the $\pi$ fields taken from $V$, and $r$ the number of derivatives acting on the $\pi$ fields coming from  $O$. One differentiates \eqref{eq:twopoint} $s$ times with respect to $y$ and $r$ times with respect to $x$, and then sets $x$ to the origin. The result of this Wick contraction can be written as a (possibly tensorial) function of $y$, multiplying an operator at the origin with $m-2$ $\pi$ fields and $p-r$ derivatives, together with an operator at position  $y$ (the remainder of the interaction) containing two $\pi$ fields and $2-s$ derivatives. 

\item Taylor expand the remaining $y$-dependent operator around the origin. Since the operator $O$ can only mix with operators of the same classical dimension, poles in $\epsilon$ can only arise from operators that contain $p$ derivatives as well. This implies that the only relevant contribution in the Taylor expansion is the one that yields  an operator at the origin with $p$ derivatives, corresponding to the term of order $(s+r-2)$. All the other terms in the expansion can be neglected. 
\item Use Lorentz invariance to reduce the remaining integral over $y$ to a scalar integral and evaluate it using
\begin{equation}\label{eq:integratexpowers}
\int_{|y|\le 1} \mathrm d^{d} y \, \frac{1}{|y|^d}
=
-\frac{1}{2\pi\epsilon}
+\text{finite}\,.
\end{equation}
\end{enumerate}
The final result is a pole in $\epsilon$ times a linear combination of operators with $m$  $\pi$ fields and $p$ derivatives. If operator mixing is present, it may be necessary to use linear algebra techniques to re-express this result in terms of the independent operators of the basis to which $O$ belongs.
\bibliographystyle{utphys}
\bibliography{NLSM}

@article{Zan:2026oyb,
	author = "Zan, Bernardo",
	title = "{On the Wilson-Fisher fixed point in the limit of integer spacetime dimensions}",
	eprint = "2601.19977",
	archivePrefix = "arXiv",
	primaryClass = "hep-th",
	month = "1",
	year = "2026"
}

@article{Henriksson:2025hwi,
    author = "Henriksson, Johan and Herzog, Franz and Kousvos, Stefanos R. and Roosmale Nepveu, Jasper",
    title = "{Multi-loop spectra in general scalar EFTs and CFTs}",
    eprint = "2507.12518",
    archivePrefix = "arXiv",
    primaryClass = "hep-ph",
    reportNumber = "CERN-TH-2025-135",
    month = "7",
    year = "2025"
}

@article{Henriksson:2025vyi,
    author = "Henriksson, Johan and Kousvos, Stefanos R. and Roosmale Nepveu, Jasper",
    title = "{EFT meets CFT: Multiloop renormalization of higher-dimensional operators in general $\phi^4$ theories}",
    eprint = "2511.16740",
    archivePrefix = "arXiv",
    primaryClass = "hep-th",
    reportNumber = "CERN-TH-2025-258",
    month = "11",
    year = "2025"
}

@book{Cardy:1996xt,
    author = "Cardy, John L.",
    title = "{Scaling and renormalization in statistical physics}",
    year = "1996"
}

@article{Hogervorst:2014rta,
	author = "Hogervorst, Matthijs and Rychkov, Slava and van Rees, Balt C.",
	title = "{Truncated conformal space approach in $d$ dimensions: A cheap alternative to lattice field theory?}",
	eprint = "1409.1581",
	archivePrefix = "arXiv",
	primaryClass = "hep-th",
	reportNumber = "CERN-PH-TH-2014-155",
	doi = "10.1103/PhysRevD.91.025005",
	journal = "Phys. Rev. D",
	volume = "91",
	pages = "025005",
	year = "2015"
}

@article{Hogervorst:2015akt,
	author = "Hogervorst, Matthijs and Rychkov, Slava and van Rees, Balt C.",
	title = "{Unitarity violation at the Wilson-Fisher fixed point in 4-$\epsilon$ dimensions}",
	eprint = "1512.00013",
	archivePrefix = "arXiv",
	primaryClass = "hep-th",
	reportNumber = "CERN-PH-TH-2015-282, YITP-SB-15-44, DCPT-15-65",
	doi = "10.1103/PhysRevD.93.125025",
	journal = "Phys. Rev. D",
	volume = "93",
	number = "12",
	pages = "125025",
	year = "2016"
}

@article{paper1,
    author = "De Cesare, Fabiana and Rychkov, Slava",
    title = "{Disturbing news about the $d=2+\epsilon$ expansion}",
    eprint = "2505.21611",
    archivePrefix = "arXiv",
    primaryClass = "hep-th",
    doi = "10.1093/ptep/ptaf103",
    journal = "PTEP",
    volume = "2025",
    number = "09",
    pages = "093B02",
    year = "2025"
}

@book{Cardy-book,
	author         = "Cardy, John L.",
	title          = "{Scaling and renormalization in statistical physics}",
	year = "1996",
	publisher        = "Cambridge, UK: Univ. Pr., 238 p.",
	SLACcitation   = "%%CITATION = INSPIRE-429658;%%"
}

@article{Henriksson:2022rnm,
    author = "Henriksson, Johan",
    title = "{The critical O(N) CFT: Methods and conformal data}",
    eprint = "2201.09520",
    archivePrefix = "arXiv",
    primaryClass = "hep-th",
    doi = "10.1016/j.physrep.2022.12.002",
    journal = "Phys. Rept.",
    volume = "1002",
    pages = "1--72",
    year = "2023"
}

@article{Nahum_2015,
   title={Deconfined Quantum Criticality, Scaling Violations, and Classical Loop Models},
   volume={5},
   DOI={10.1103/physrevx.5.041048},
   number={4},
   journal={Physical Review X},
   author={Nahum, Adam and Chalker, J. T. and Serna, P. and Ortuño, M. and Somoza, A. M.},
   year={2015},
    archivePrefix = {arXiv},
   eprint = {1506.06798},
   primaryClass = {cond-mat.str-el} }

@article{Polyakov:1975rr,
    author = "Polyakov, Alexander M.",
    title = "{Interaction of Goldstone Particles in Two Dimensions. Applications to Ferromagnets and Massive Yang-Mills Fields}",
    doi = "10.1016/0370-2693(75)90161-6",
    journal = "Phys. Lett. B",
    volume = "59",
    pages = "79--81",
    year = "1975"
}

@phdthesis{Jones:2024ept,
	author = "Jones, Robert A.",
	title = "{Explorations in two dimensional strongly correlated quantum matter: from exactly solvable models to conformal bootstrap}",
	school = "MIT",
	year = "2024",
	supervisor = "Max A. Metlitski"
}

@article{Brezin:1975sq,
    author = "Brezin, E. and Zinn-Justin, Jean",
    title = "{Renormalization of the nonlinear sigma model in $2 +\epsilon$ dimensions. Application to the Heisenberg ferromagnets}",
    doi = "10.1103/PhysRevLett.36.691",
    journal = "Phys. Rev. Lett.",
    volume = "36",
    pages = "691--694",
    year = "1976"
}

@article{Bardeen:1976zh,
    author = "Bardeen, William A. and Lee, Benjamin W. and Shrock, Robert E.",
    title = "{Phase Transition in the Nonlinear $\sigma$ Model in $2 + \epsilon$ Dimensional Continuum}",
    reportNumber = "FERMILAB-PUB-76-033-THY, FERMILAB-PUB-76-033-T",
    doi = "10.1103/PhysRevD.14.985",
    journal = "Phys. Rev. D",
    volume = "14",
    pages = "985",
    year = "1976"
}

@article{Brezin:1976ap,
    author = "Brezin, E. and Zinn-Justin, Jean and Le Guillou, J. C.",
    title = "{Renormalization of the Nonlinear Sigma Model in $2 + \epsilon$ Dimension}",
    reportNumber = "SACLAY-DPH-T-76-58",
    doi = "10.1103/PhysRevD.14.2615",
    journal = "Phys. Rev. D",
    volume = "14",
    pages = "2615",
    year = "1976"
}

@article{Brezin:1976qa,
    author = "Brezin, E. and Zinn-Justin, Jean",
    title = "{Spontaneous Breakdown of Continuous Symmetries Near Two-Dimensions}",
    reportNumber = "SACLAY-DPh.T/76/27",
    doi = "10.1103/PhysRevB.14.3110",
    journal = "Phys. Rev. B",
    volume = "14",
    pages = "3110",
    year = "1976"
}

@article{Hikami:1977vr,
    author = "Hikami, S. and Brezin, E",
    title = "{Three Loop Calculations in the Two-Dimensional Nonlinear Sigma Model}",
    reportNumber = "SACLAY-DPh-T-77/131",
    doi = "10.1088/0305-4470/11/6/015",
    journal = "J. Phys. A",
    volume = "11",
    pages = "1141--1150",
    year = "1978"
}

\end{document}